\documentclass[aps,pre,twocolumn,superscriptaddress]{revtex4}

\usepackage{amsfonts,amssymb,amsmath,latexsym,epsfig} 

\begin{document}

\title{Poincar\'e recurrences and transient chaos in systems with leaks}

\author{Eduardo G. Altmann}
\affiliation{Max Planck Institute for the Physics of Complex Systems,  01187 Dresden, Germany}
\affiliation{Northwestern Institute on Complex Systems, Northwestern University, Evanston, Illinois 60208, USA}

\author{Tam\'as T\'el}
\affiliation{Institute for Theoretical Physics, E\"otv\"os University, P.O. Box 32, H-1518 Budapest, Hungary}

\begin{abstract}

In order to simulate observational and experimental situations, we consider a leak in
the phase space of a chaotic dynamical system. 
We obtain an expression for the escape rate of the survival probability applying the
theory of transient chaos.
This expression improves previous estimates based on the properties of the closed system and 
explains dependencies  on the position and size of the leak and on the initial ensemble. 
With a subtle choice of the initial ensemble, we obtain an equivalence to the classical
problem of Poincar\'e recurrences in closed systems, which is treated in the same
framework. Finally, we show how our results apply to weakly  
chaotic systems and justify a split of the invariant saddle in  hyperbolic and nonhyperbolic components,
related, respectively, to the intermediate exponential and asymptotic power-law decays of the survival probability.
\end{abstract}

\maketitle


\section{Introduction}\label{sec.intro}
The numerical and theoretical study of dynamical systems can resemble
observational and experimental situations if the escape of an
initial ensemble of trajectories through a leak placed in the otherwise 
closed phase space is considered.
The idea of introducing a leak in a closed chaotic dynamical system to generate transient chaos was 
first suggested 
by Pianigiani and Yorke \cite{PY}. Later this
problem was discussed in detail in the context of fractal exit
boundaries \cite{BGOB}, of geometrical acoustics \cite{LS}, 
of quantum chaos \cite{DS}, of controlling chaos \cite{paar},  
of resetting in hydrodynamical flows \cite{pierrehumbert,schneider}, 
of leaked Hamiltonian systems \cite{SHA}, 
of astronomy \cite{nagler.astro} and 
of cosmology \cite{motter}.
The subject has recently received a renewed attention
\cite{optics,altmann.optics,kuhl,nagler.billiards,PCV,BD,BY}. Part of this renewed
interest comes from recent developments of quantum chaos, that started 
to consider carefully the effect of measurement devices, absorption,
and other forms of leaking both theoretically and
experimentally~\cite{kuhl,altmann.optics,optics,nagler.billiards}. 
It is by now clear that leaking dynamical systems provides a way 
to 'peeping at chaos' \cite{BD}, and can be used as a kind
of chaotic spectroscopy \cite{DS}.

From a theoretical point of view, leaking systems provide an interesting bridge between open and closed dynamical
systems. In closed systems the phase space coordinates remain confined,
and a rigorous mathematical approach based on attractors and ergodic
components can be employed \cite{ott}. 
An important quantity is the distribution
$p_r(T)$ of the first Poincar\'e recurrence times~$T$ 
\cite{kac,dorfman,Zasl1,ZT,Zasl2,CS} to a preselected region of the phase space.
As originally proposed by Chirikov and Shepelyansky \cite{CS} (see also \cite{chirikov}), $p_r(T)$
is a useful analyser of the entire dynamics.  
In contrast, in open systems trajectories may leave the region of interest (e.g., tend to $\pm \infty$)
and the method of transient chaos is usually employed \cite{ott,TG}. In this case, the dynamics is characterized
by the escape time distribution~$p_e(n)$ 
(the derivative of the survival probability). 

Chaotic dynamics usually leads to an asymptotic exponential decay of 
both $p_r$ and $p_e$ 
\begin{equation}\label{eq.exp}
p_{r,e}(t) \sim e^{-\gamma_{r,e} t},
\end{equation}
for large enough $t=\{T,n\}$. Exponentials are the signature of {\em strong} chaotic
properties as seen, e.g., in the divergence of nearby initial conditions,
in the decay  of correlations, and in the convergence to equilibrium
distributions.  The exponent~$\gamma_{r,e}$ is the key quantifier of system
specific characteristics and will be considered in detail in this paper.

The possibility of comparing a system with leak to its corresponding closed system 
shows numerous advantages in comparison to naturally opened
systems. For instance, the rate~$\gamma_{r,e}$ in Eq.~(\ref{eq.exp}) can be
estimated from the probability of escaping or returning. For small leaks, 
this probability
is obtained by computing the closed systems natural
measure~$\mu$  of a typical leak region $I$, and it follows
that~\cite{paar,Zasl2,altmann}
\begin{equation}\label{eq.smallmu}
\gamma_r=\gamma_e=\mu(I)= \frac{1}{\langle t \rangle} \;\; \text{ for } \;\; 1 \gg \mu(I) > 0,
\end{equation}
where $\langle t \rangle$ denotes the mean recurrence or escape times, $\langle
T \rangle$ or $\langle n \rangle$.

Despite its simplicity, relation~(\ref{eq.smallmu}) is of little practical 
relevance since the
limit $\mu(I)\rightarrow0$ is never achieved in either numerical or
experimental applications~\cite{altmann,BKG,maths2}. Here we do not restrict
ourselves to this limit and find that, apart from being unrealistic, it masks
  different interesting relations.  
We perform a formal description through the theory of
transient chaos, i.e., in terms of a 
nonattracting chaotic set \cite{ott} and the conditionally
invariant measure~\cite{PY,T},  and  obtain distinct relations for~$\gamma$ and  
  $1/\langle t \rangle$ as a function of the  position and size of the
  region~$I$ and of the initial ensemble.  Relation~(\ref{eq.smallmu}) does
  not hold, but we show that $\gamma_r=\gamma_e\equiv \gamma$ remains
  valid. This is based on a full correspondence with the recurrence problem
  obtained for a special initial ensemble and setting the 
 {\em leak to be equal to the recurrence region of the closed system}. We have explored 
 the most striking consequences of this unified treatment from the point of view of
 Poincar\'e recurrences in a previous short paper~\cite{letter}. Here, after describing 
the relationship between the two problems with additional details, we focus on the escape problem and
 consider the recurrence problem as a particular, yet important, case. We also  
 use this connection to apply and adapt a method to  determine~$\gamma$~\cite{altmann} based only
on the escape probabilities for {\em short} times. This is a main advantage
over genuine open systems where long-term periodic orbits are needed to
determine~$\gamma_e$. 

In systems showing {\em weak} chaos, as e.g., 
Hamiltonian systems with mixed phase space,
the exponential decay law~(\ref{eq.exp}) experiences a cross-over for longer
times towards an asymptotic power-law behavior~\cite{chirikov,Zasl2,JTZ}. We argue,
however, that  on intermediate times a decay  rate~$\gamma$ can be
well defined  
\begin{equation}\label{eq.powerlaw}
p_{r,e}(t) \propto \left\{ \begin{array}{ll} e^{-\gamma t} & \text{ for
  intermediate} \;\; t < t_c, \\
  t^{-\alpha} & \text{ for } t > t_c. \\
\end{array} 
\right.
\end{equation}
We interpret our results based on an effective splitting of the chaotic
saddle into hyperbolic and nonhyperbolic components.

The paper is organized as follows. In Sec.~\ref{sec.recurrence} we 
review the main properties of  Poincar\'e recurrence
times. The corresponding results for escape times appear in
Sec.~\ref{sec.escape}, together with the dependence of $\gamma$ and~$\langle t \rangle$ on~$I$. The
equivalence between recurrence and escape problems
is carefully discussed in Sec.~\ref{sec.rec}. In Sec.~\ref{sec.henon} numerical illustrations of
the results are provided for paradigmatic dissipative models. 
In Sec.~\ref{sec.standard}, the nonhyperbolic case is investigated using an

area-preserving map. Relations to periodic orbits and an efficient method to
determine~$\gamma$ are described in Sec.~\ref{sec.fitting}. Finally,
conclusions and discussions appear in Sec.~\ref{sec.conclusion},
where the case of higher order recurrences is also investigated.


\section{Poincar\'e recurrences in closed systems}\label{sec.recurrence}

Poincar\'e recurrences to a certain region
of the phase space have played an important role 
since the foundation of the kinetic description of noneqilibrium processes \cite{Zasl1,Zasl2}. 
We define the problem in the context of low-dimensional chaotic maps that can
be dissipative~\cite{gao} or Hamiltonian~\cite{CS,Zasl2}.

Consider a discrete time chaotic system $\vec{x}_{n+1}=M(\vec{x}_n)$ defined
on a bounded phase space $\vec{x}\in\Gamma$. Let the natural ergodic measure
on the invariant chaotic set (e.g., strange attractor or chaotic sea) be
denoted by~$\mu$ and its density by~$\rho_\mu$. We define the {\em recurrence
  region}  as a subset $I \subset \Gamma$ with $\mu(I)>0$. As stated above, we {do {\em
    not} restrict ourselves to the unrealistic limit
  $\mu(I)\rightarrow0$}~\cite{altmann,BKG,maths2}. The Poincar\'e 
recurrence theorem ensures that for almost   
all initial conditions~$\vec{x}_0\in I$, there are infinitely many
time instants $n=n_1, n_2, \ldots$ such that $M^n(\vec{x}_0) \in I$. The
first recurrence times are defined as $T_i=n_i-n_{i-1}$, with 
$i\geq 1$  and $n_0\equiv 0$  (if the points remains in $I$ at the $i$th iterate $T_i=1$).
The recurrence time distribution~$p_r(T)$, $T \ge 1$, is 
the probability of finding $T_i\equiv T$ 
in an infinitely long trajectory.
The cumulative version can be written as $P_r(\tau)=\sum_{T=\tau}^\infty p_r(T)$. The mean recurrence time is defined as 
$$\langle T \rangle \equiv \lim_{N\rightarrow\infty}\frac{1}{N}\sum_{i=1}^{N}
T_i \equiv \sum_{T=1}^{\infty} T p_r(T).$$
Due to the ergodicity of the measure~$\mu$, the specific choice of the initial
point~$\vec{x}_0 \in I$ is irrelevant for $p_r(T)$. 
Instead of following a single trajectory, $p_r(T)$ can also be
obtained by using an ensemble of trajectories
chosen inside~$I$ according to 
the invariant density~$\rho_\mu$. A consequence of ergodicity is Kac's lemma \cite{kac}:
\begin{equation}\label{eq.kac}
\langle T \rangle = \frac{1}{\mu(I)},
\end{equation}
which is valid for any recurrence region~$I$.

The recurrence time distribution of an uncorrelated  random process with fixed return
  probability ~$\mu$ follows a binomial distribution~\cite{altmann}
\begin{equation}\label{eq.binomial}
p_r(T) = \mu (1-\mu)^{T-1}=\frac{\mu}{1-\mu} e^{\ln(1-\mu) T}.
\end{equation}
The relaxation rate~$\gamma_r$ defined in Eq.~(\ref{eq.exp}) is thus given
for such a random process by
\begin{equation}\label{eq.gammamu}
  \gamma^*_r=-\ln(1-\mu).
\end{equation} 
For small recurrence probability $ \mu \rightarrow 0$, $\gamma^*_r \rightarrow \mu$, and the Poisson
  distribution~$p_r(T)=\mu \exp(-\mu  T)$ is obtained. 
Identifying $\mu$ with $\mu(I)$ (recurrence probability equals to the natural
measure of the recurrence region), we see that a Poisson process is the simplest process
  leading to Eq.~(\ref{eq.smallmu}). Indeed, the Poisson distribution is proved to
  describe deterministic hyperbolic systems (see references in~\cite{maths}).

We study next the distribution of recurrence times $p_r(T)$ for generic
chaotic systems. In agreement with Eq.~(\ref{eq.exp}), we write 
\begin{equation}\label{eq.prec}
p_r(T) \approx \left\{ \begin{array}{ll} \text{ irregular }& \text{ for } 1 < T < T^*, \\
   g_r e^{-\gamma_r T} & \text{ for } T \geq T^*. \\
                               \end{array} 
\right.
\end{equation}
Implicit is the view that the approach toward the exponential distribution
for $T\rightarrow \infty$ does not occur uniformly, but that 
after some short time~$T^*$ the distribution is practically very close to an
exponential~\cite{altmann}. This will be theoretically justified in the next section and
numerically illustrated in Sec.~\ref{sec.henon}. 
We are interested in the decay rate~$\gamma_r$, which typically 
deviates from the random estimate  (\ref{eq.gammamu}), and also in the short time
irregular fluctuations for~$T<T^*$, where~$T^*$ is of the order of 
$\langle T \rangle$ and will be better interpreted in
Sec.~\ref{sec.rec}. These short time irregular oscillations [denoted as {\em irregular} in
Eq.~(\ref{eq.prec}) and equations hereafter] depend strongly on the 
choice of the recurrence region~$I$. 
For instance, if a periodic orbit of (short)
period $p$ is present inside~$I$, there is a higher probability of recurrence
at~$T=p$. 
As shown in Ref.~\cite{altmann} and will be discussed in
Sec.~\ref{sec.fitting}, normalization and 
Eq.~(\ref{eq.kac}) imply that the short-time fluctuations 
uniquely determine
the deviation of $\gamma_r$ from $\gamma^*_r$.   
Therefore, for different recurrence regions~$I$ of the same natural
measure~$\mu(I)$ [while $\langle T \rangle$ is fixed and given
by~(\ref{eq.kac})], the value of~$\gamma_r$ depends strongly on the
position of~$I$~\cite{altmann}.


\section{Escape in open systems}\label{sec.escape}

\subsection{General relations for transiently chaotic systems}\label{ssec.general}

By now, transient chaos has a well established theory \cite{ott,TG}. 
In this case, the discrete system~$\vec{x}_{n+1}=\tilde{M}(x_n)$ is defined in
an unbounded phase space~$\tilde{\Gamma}$.  
A region of interest $A\subset \tilde{\Gamma}$ 
(non-trivial dynamics) is defined in such a way that trajectories
may leave~$A$ [$\tilde{M}(A) \nsubseteq A$] but do not return to it
[$\tilde{M}(\tilde{\Gamma} \setminus A) \cap A \equiv \varnothing$]. An
initial ensemble is started
according to a smooth density~$\rho_0(\vec{x})$, with $\vec{x} \in 
A$. The escape time distribution $p_e(n)$ is given by
the fraction of trajectories that leave~$A$ at time~$n\geq1$. For chaotic
systems it can be written as 
\begin{equation}\label{eq.pesc}
p_e(n) = \left\{ \begin{array}{ll} \text{ irregular }& \text{ for } 1 < n < n^*_e, \\
   g_e e^{-\gamma_e n} & \text{ for } n \geq n^*_e, \\
                               \end{array} 
\right.
\end{equation}
where $n^*_e$ corresponds to a short convergence time and $\gamma_e$ is the escape
rate. 
The survival probability inside $A$ up to time $n$ is given by $P_e(n)=\sum_{n+1}^\infty p_e(n')$, and is also an exponential distribution with the same $\gamma_e$ as in Eq.~(\ref{eq.pesc}). The
average escape time, $\langle n \rangle_e\equiv\sum_1^\infty n p_e(n)$, also called the lifetime of chaos, is usually estimated by the reciprocal of $\gamma_e$ \cite{ott,TG}. 
Note, however, that the exact average is, in general, different from $1/\gamma_e$ and {\em depends} (just like $n_e^*$) on $\rho_0$.

The ergodic theory of transient chaos explains the above properties by the existence of
a non-attracting chaotic set. In invertible systems this set is, e.g., a chaotic 
saddle\footnote{In noninvertible systems the set is a chaotic repeller.}: the set of
  orbits that never escape  either forward or backward in
  time~\cite{dorfman,ott,TG}. For systems with strong chaos, the saddle is an invariant fractal set
  of zero measure. Long-time transients correspond to trajectories that
  approach it closely, along its stable manifold, and leave it along its
  unstable manifold.   Being an invariant object, the chaotic saddle is, of course, 
independent of the initial density $\rho_0$.
Therefore, assuming that the distribution~$\rho_0$ is non-vanishing
  in at least some region around the stable manifold of the saddle (and that
  there are no disjoint saddles) the escape 
  rate~$\gamma_e$ is {\em independent} of the initial density. It can be related to
the geometrical and dynamical  properties of  the saddle by the Kantz-Grassberger
formulas~\cite{kantzgrassberger}, which for two-dimensional invertible maps are
\begin{equation}\label{eq.kantzgrassberger}
\gamma_e=\lambda (1-d_u), \;\;\;\; \lambda d_u=- \lambda' d_s,
\end{equation}
where $\lambda$ ($\lambda'$) is the positive (negative) Lyapunov exponent on the saddle and $d_u$ ($d_s$) is the
information dimension of the saddle along its unstable (stable) manifold
\cite{ott,TG}. The time
$n^*_e$, below which  orbits escape without approaching the saddle, can be
estimated by the inverse of the negative Lyapunov exponent $\lambda'$.

A rigorous description of open systems is possible in terms
of the {\em conditionally invariant measure}~\cite{PY,T,DY,chernov}, hereafter called {\em c-measure}. A
measure~$\mu_c$ is said to be conditionally invariant if
\begin{equation}\label{eq.conditional}
\frac{\mu_c(\tilde{M}^{-1}(E))}{\mu_c(\tilde{M}^{-1}(A))}=\mu_c(E),
\end{equation}
for any $E \subset A$. In words: the c-measure is not invariant under the map:
${\mu_c(\tilde{M}^{-1}(E))} \neq \mu_c(E)$, but is preserved under the incorporation of the
compensation factor in the denominator of (\ref{eq.conditional}). The
compensation factor $\mu_c(\tilde{M}^{-1}(A))<1$ corresponds to the c-measure
of the set remaining inside~$A$, i.e., the trajectories that do not escape in one
iteration.
The c-measure of the region of interest is $\mu_c(A)=1$.
The c-measure has a density $\rho_c$, which is
unique and the only attractor for smooth initial densities $\rho_0$. Taking
into account the exponential escape expressed
in~(\ref{eq.pesc}), one sees that  convergence is only possible if \cite{PY}
\begin{equation}\label{eq.gammageneral}
\gamma_e=-\ln(\mu_c(\tilde{M}^{-1}(A))).
\end{equation}
Considering the evolution of $\rho_0$ through an evolution operator, the
Perron-Frobenius operator~\cite{dorfman,T}, $\exp(\gamma_e)$ is the largest eigenvalue and
$\rho_c$ is the corresponding (right) eigenvector. 
The c-measure describes the escape process of the system and therefore~$\rho_c$ is concentrated on the saddle and along its unstable manifold~\cite{T}.

Numerically, $\rho_c$ is obtained multiplying $\rho_0$ at each iteration by a
constant factor which, after many iteration, is necessarily equal to $\exp(\gamma_c)$. In
practice, we iterate a large number of trajectories and we obtain $\rho_c$ 
by renormalizing the surviving trajectories at a long
time. $\mu_c(\tilde{M}^{-1}(A))$ in~(\ref{eq.gammageneral}) [or $1-\mu_c(I)$
in~(\ref{eq.gammamuc}) below] is obtained as the fraction of all surviving trajectories
that escape in the next time step or, equivalently, by fitting $\gamma_e$ to~$p_e(n)$ and inverting~(\ref{eq.gammageneral}).

\subsection{Systems with leaks}

Open systems~$\tilde{M}$ can be obtained from the closed system~$M$ defined on $\Gamma$, described
in Sec.~\ref{sec.recurrence}, by introducing a leak in a region~$I \subset
\Gamma$   
\begin{equation}\label{eq.leak}
\vec{x}_{n+1}=\tilde{M}(\vec{x}_n) = \left \{ \begin{array}{ll} M(\vec{x}_n) & \text{ if } \vec{x}_n \notin I \\
                                       \text{ escape } & \text{ if } x_n \in I .\\
                                       \end{array} \right.
\end{equation}
Notice that, since the escape happens one step {\em after} entering~$I$, the map $\tilde{M}$ is defined in~$I$ and initial conditions can be in $I$.
This procedure defines an open system as that of Sec.~\ref{ssec.general}, with $A \equiv \Gamma$. 

An estimate~$\gamma^*_e$ of the escape rate~$\gamma_e$ in systems with leaks is usually given in terms
of the Frobenius-Perron relation~\cite{paar,dorfman} which expresses that
the number of particles not escaping within one time step is
proportional  to the natural measure 
outside of the leak: $\exp{(-\gamma^*_e)}=1-\mu(I)$, i.e., 
\begin{equation}\label{eq.fpe}
\gamma^*_e=-\ln(1-\mu(I)) \;\; [\approx \mu(I) \text{ for } \;\; \mu(I)\rightarrow 0],
\end{equation}
where $\mu(I)$ is the natural measure of leak $I$.
This relation is equivalent to the binomial estimate for recurrence
times~(\ref{eq.gammamu}) and we call it hereafter as the
naive estimate of~$\gamma_e$. Deviations of $\gamma_e$ from $\gamma^*_e$ were reported in
Refs.~\cite{paar,schneider} and associated to the existence of short-time periodic orbits
of the closed system inside~$I$.  

The results of Sec.~\ref{ssec.general} on open systems can be applied to systems with leaks. Since $\tilde{M}^{-1}(A)\equiv A \setminus I$, it
follows that $\mu_c(\tilde{M}^{-1}(A))=\mu_c(A)-\mu_c(I)=1-\mu_c(I)$, and the
following simple relation is obtained from~(\ref{eq.gammageneral})
\begin{equation}\label{eq.gammamuc}
\gamma_e = -\ln(1-\mu_c(I)).
\end{equation}
This exact expression, which was previously obtained in Ref.~\cite{paar}, corresponds to the naive estimate~(\ref{eq.fpe})
replacing~$\mu(I)$ by $\mu_c(I)$. They become equivalent
in the limit of small leak region~$\mu(I)\rightarrow 0$ because $\mu_c$ tends to~$\mu$ \cite{ML,AS}.
It is the c-measure of the {\em leak} which determines the escape rate. 
Even though numerically such a measure is easily calculated. Proofing the existence of
such conditionally invariant measure may be a very 
involving mathematical task~\cite{PY,DY} (see, e.g., Ref.~\cite{chernov} for a class of
maps with leaks).
Here we employ a pragmatic approach and implicitly
assume that such a measure exists in the large class of systems with leaks where numerical simulations indicate exponential decay of $p_e(n)$.

\subsection{Dependence of the mean escape time on the initial density}\label{ssec.ic}

While the escape rate~$\gamma_e$ is independent of the initial density $\rho_0$, as expressed by Eq.~(\ref{eq.gammamuc}), the mean
escape time~$\langle n \rangle_e$ and the full distribution~$p_e(n)$ 
change considerably with $\rho_0$. It is thus instructive to discuss the following 
special initial conditions:

{\em Conditionally invariant density $\rho_0=\rho_c$:} 
from a formal mathematical perspective the natural choice of initial
conditions is according to the density of the c-measure~$\rho_c(\vec{x})$, 
which can be considered as the natural measure of the open system. 
This corresponds to setting
the initial density in agreement with the escape process. Therefore, no
oscillations are observed in Eq.~(\ref{eq.pesc}), i.e., $n^*=1$. Since~$\gamma_e$ is given by~(\ref{eq.gammamuc}) and the distribution is normalizable, $p_e(n)$ follows a binomial distribution 
$$p_e(n) = \mu_c (1-\mu_c)^{n-1}=\frac{\mu_c}{1-\mu_c} e^{\ln(1-\mu_c) n},$$
with $\mu_c=\mu_c(I)$, in analogy with Eq.~(\ref{eq.binomial}). The mean
escape time is in this case
\begin{equation}\label{eq.kac2}
\langle n \rangle_c = \frac{1}{\mu_c(I)}=\frac{1}{1-e^{-\gamma_e}}, \;\;\;\;
\;\;\; \text{with} \;\;\; \rho_0=\rho_c. 
\end{equation}

{\em Natural density $\rho_0=\rho_\mu$:} 
coherent with the idea of systems with leaks, the initial density is chosen according to the natural measure of the closed system over the full $\Gamma$.
It corresponds to leaking the system after equilibrium has been achieved in the closed 
system. We show in Sec.~\ref{sec.henon} that 
\begin{equation}\label{eq.kac3}
\langle n \rangle_\mu \approx \frac{1}{\mu_c(I)}=\frac{1}{1-e^{-\gamma_e}} \;\;\;  
\text{with}, \;\;\; \rho_0=\rho_\mu. 
\end{equation} 
Similar results are obtained for densities proportional to~$\rho_\mu$ (equal to $\rho_\mu$
apart from normalization in a subspace of $\Gamma$).

{\em Smooth density $\rho_0=\rho_s$:} 
in some systems, and for
numerical simulations, it is natural to define the initial
density to be a smooth  function over $\Gamma$ or a part of $\Gamma$. 
A particular case is that of a uniform density.
Estimate (\ref{eq.kac3}) remains valid for such initial densities as well, as
shown in Sec.~\ref{sec.henon}.

\section{Relation between recurrence and escape}
\label{sec.rec}

In this section we show that there is an initial density $\rho_0=\rho_r$ for
which $p_r(T)=p_e(n)$ for $T=n$. We use 
this special initial condition to establish a relation between recurrence and escape. 

\begin{figure}[!ht]
\includegraphics[width=0.9\columnwidth]{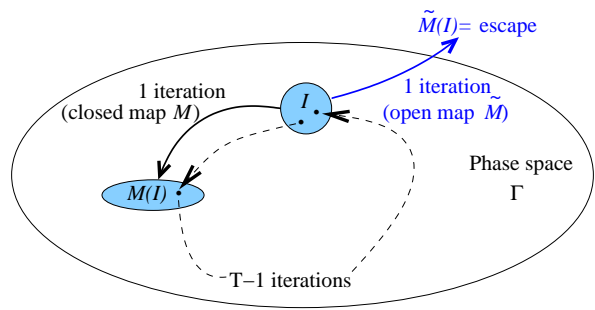}
\caption{ (Color online) Illustration of the equivalence between recurrence and
  escape times.}
\label{fig.ilust}
\end{figure}

\begin{table*}
\begin{center}
\begin{tabular}{|l|c c c|c|}
\hline
{}
&{}
&{Large leaks}
&{}
&{Small leaks}
\\\hline
{Measures}
&{}
&{Finite $\mu(I)\neq\mu_c(I)$}
&{}
&{$\mu(I)=\mu_c(I) \rightarrow 0$}
\\\hline
{Escape rate}
&{}
&{$\gamma=-\ln{(1-\mu_c(I))}\neq -\ln(1-\mu(I))=\gamma^*$}
&{}
&{$\gamma=\mu_c(I) =\mu(I)$}
\\\hline
{Mean  time}
&{$\langle n \rangle_r=\dfrac{1}{\mu(I)}\neq \dfrac{1}{1-e^{-\gamma}}$}
&{$\langle n \rangle_c=\dfrac{1}{\mu_c(I)}=\dfrac{1}{1-e^{-\gamma}}$}
&{$\langle n \rangle_{\mu,s} \approx\dfrac{1}{\mu_c(I)}=\dfrac{1}{1-e^{-\gamma}}$}
&{$\langle n \rangle =\dfrac{1}{\mu(I)}=\dfrac{1}{\gamma}$}
\\
{~$\rho_0$}
&{ Recurrence: $\rho_r$}
&{  c-measure: $\rho_c$}
&{ natural, smooth: $\rho_{\mu,s}$}
&{$\rho_{r,c,\mu,s}$}
\\\hline
\end{tabular}
\caption{ (Color Online) The escape/recurrence rate~$\gamma=\gamma_r=\gamma_e$ and the mean escape time (recurrence time for initial density $\rho_r$)~$\langle n \rangle$ expressed
  in terms of the natural invariant measure~$\mu(I)$ and the conditionally invariant
  measure~$\mu_c(I)$ of the leak/recurrence region~$I$, for large and small
  leaks.}\label{tab.ic} 
\end{center}
\end{table*}


{\em Density equivalent to recurrence $\rho_0=\rho_r$:} consider that 
trajectories are injected from outside through the leak. More precisely, consider the
distribution~$\rho_r$ obtained as the  
first iterate through map~$M(\vec{x})$ of the points $\vec{x}\in I$ distributed according
to the natural density $\rho_\mu$. This can be obtained applying the Perron-Frobenius
operator~\cite{dorfman} as 
\begin{equation}\label{eq.rhor1}
\rho_r(\vec{x}) = \frac{\rho_\mu(M^{-1}(\vec{x})\cap I)}{J(M^{-1}(\vec{x})\cap I)\mu(I)} \;\; \text{ for } \;\; \vec{x}\in M(I),
\end{equation}
where~$M^{-1}(x)\cap I$ denotes the points that come from $I$, $J$ is the Jacobian of the
map, and $\mu(I)$ ensures normalization~\footnote{This extends the validity of Eq.~(5) of
  Ref.~\cite{letter} to maps with non-constant J.}.  
In the case of invertible maps, $M^{-1}$ is unique and the density~(\ref{eq.rhor1}) is equivalent to
\begin{equation}\label{eq.rhor2}
\rho_{r}(\vec{x})=  \left\{  \begin{array}{ll}
      {\frac{\rho_{\mu}(M^{-1}(\vec{x}))}{J(M^{-1}(\vec{x}))\mu(I)}} & \text{ if } \; \vec{x} \in M(I), \\
      0 & \text{ else}.\\
\end{array}
\right.
\end{equation}
Consider now an infinitely long trajectory used to calculate the recurrence
times~$T$ described in Sec.~\ref{sec.recurrence}. In view of the ergodic theorem
(time average equals  ensemble average), we see that {\em one iteration after} returning to~$I$
the points of this
trajectory are distributed
precisely as~$\rho_r$. As illustrated in Fig.~\ref{fig.ilust}, due to the one step time shift inserted in
definition~(\ref{eq.leak}), all escape times~$n$ of~$\rho_r$ correspond to a
recurrence time~$T$\footnote{Escapes at $n=1$ occur when $I\cap M(I) 
\neq 0$ and $p_e(1)=\mu(I \cap M(I))$.}. Since this is valid for all~$n$,
$p_{e}(n) \equiv p_r(T)$ for $\rho_0=\rho_r$.  In   
particular,  $\gamma_r=\gamma_e$, and since $\gamma_e$ is independent of the initial density, this implies that
for a fixed recurrence/escape region~$I$ the decay rate of the Poincar\'e recurrences and 
the escape rate of the corresponding leaked system coincide: 
\begin{equation}\label{eq.gamma}
\gamma_r = \gamma_e \equiv \gamma. 
\end{equation}
The mean recurrence time is given, however, according to Kac's
lemma~(\ref{eq.kac}) as
\begin{equation}\label{eq.kace}
\langle n \rangle_r = \langle T \rangle =\frac{1}{\mu(I)} \;\;\;\; \text{ with
} \rho_0=\rho_r.
\end{equation}
The mean recurrence time is thus determined by the
natural measure~$\mu(I)$ and
deviates considerably from the typical mean escape time given by~(\ref{eq.kac3}). The reason is that~$\rho_r$ in (\ref{eq.rhor1}) is very {\em atypical} from the point of view of the
c-measure. In fact, $\rho_c$ is concentrated along the saddle's unstable
manifold, i.e., points~$\vec{x}$ such that $\tilde{M}^{-i}(\vec{x})$ never
escape. {On the other hand,} all points on~$\rho_r$ came from the leak 
{and therefore} the support of $\rho_r$ does not overlap with the unstable  
manifold of the saddle (but it does with the stable manifold).

In view of the results above, we consider hereafter the
recurrence problem as an escape problem with $\rho_0=\rho_r$~\footnote{This is not valid if one is interested in 
  the full sequence of recurrence times~$\{T_1,
  T_2, \ldots, T_N\}$.}. Coherently, we
omit the indices $r$ and $e$, and $T$ will also be denoted  by~$n$. 
The values of the escape rate~$\gamma$ and of the mean escape time~$\langle n
\rangle$ for the different initial densities~$\rho$ are summarized in
Table~\ref{tab.ic}. 

It is now possible to compare the results of Ref.~\cite{paar} for escapes with
those of  Ref.~\cite{altmann} for recurrence. In both cases the fully chaotic
logistic map was carefully investigated and deviations of~$\gamma$ from the 
naive
estimate~$\gamma^*$ were reported. The deviations were associated with the
presence of periodic orbits inside~$I$ and interpreted in terms of the overlap
of the pre-images of~$I$ (in Ref.~\cite{paar}) and of the short time
oscillations (in Ref.~\cite{altmann}). In our unified perspective, results of
Refs.~\cite{altmann} and~\cite{paar} correspond to choosing initial
densities $\rho_r$ and  $\rho_{\mu}$ or $\rho_s$, respectively (see
Table~\ref{tab.ic}). Since the escape rate is independent of this choice, we see that both references are giving  intuitive interpretations for
the deviation of the actual~$\gamma$ from the naive estimate~$\gamma^*$.
Analogous observations in Hamiltonian systems have been reported in \cite{ZT} (recurrence)
and in \cite{schneider} (escape). In a different approach, $\gamma$ has been related to the
diffusion coefficient in both recurrence~\cite{chirikov} and escape~\cite{GN} problems.

Formally the problem can be described in terms of the c-measure and the
convergence to it. The difference between~$\gamma$ and~$\gamma^*$ correspond
to the difference between~$\mu(I)$ and~$\mu_c(I)$, as given by Eqs.~(\ref{eq.fpe})
and~(\ref{eq.gammamuc}). As emphasized in our previous
publication~\cite{letter}, this is surprising since a measure of open systems
is employed to describe properties of closed 
systems (recurrence).
In practice,
initial densities are most often taken proportionally  
to the natural density~$\rho_\mu$ or to a smooth density $\rho_s$. In 
such cases we expect a fast convergence to~$\rho_c$, and this is why 
$1/\mu_c(I)$ [Eq.~(\ref{eq.kac2})] is a better estimate of $\langle n
\rangle$ [in Eq.~(\ref{eq.kac3})] than $1/\mu(I)$ [Eq.~(\ref{eq.kac}), which is obtained by the { atypical} $\rho_0=\rho_r$]. 
The escape rate is related to $\mu_c(I)$ which is itself specially sensitive 
to the choice of the location, shape, and size of the finite leak~$I$. It is
natural to expect that~$\mu_c(I)<\mu(I)$  
since the saddle exists only
in~$A\setminus I$ and therefore $\mu_c$ is concentrated only along its unstable
manifold. This implies that the mean recurrence time~$\langle T \rangle$,
i.e., $\langle n \rangle$ with $\rho_0=\rho_r$ should be shorter than the mean
escape time~$\langle n \rangle$ for a typical $\rho_0$. This statement is
rigorously proved in Ref.~\cite{abadi} for $\rho_0=\rho_\mu$.


\section{Numerical results}\label{sec.henon}
\begin{figure}[!bt]
\includegraphics[width=0.8\columnwidth]{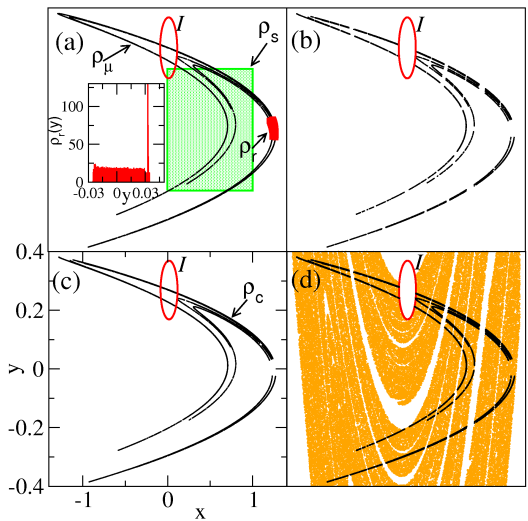}
\caption{ (Color online) H\'enon map~(\ref{eq.henon}) with a leak
  at~$(x_c,y_c)=(1.0804,0.0916)$ of radius
  $\delta=0.1$, depicted as a red circle. (a) Different initial densities: $\rho_\mu$ of the natural measure
  on the H\'enon attractor, $\rho_s =$ const. on $0<x<1,-0.2<y<0.2$, and $\rho_r$, 
  equivalent to recurrence. The inset shows the  
  nontrivial dependence of the projection $\rho_r(y)$ of $\rho_r$ on the
$y$ axis. 
(b) Invariant saddle composed of the points that do not enter $I$ for forward and backward
iterations. (c) Unstable manifold of the invariant saddle ($\rho_c$) and (d) stable
manifold in grey (orange) of the invariant saddle. 
}
\label{fig.henon}
\end{figure}

We illustrate the previous results through numerical
simulations of the dissipative H\'enon map
\begin{equation}\label{eq.henon}
x_{n+1}=1-1.4 x_n^2+y_n, \;\;\;\;
y_{n+1}=0.3 x_{n}. 
\end{equation}
Initial conditions inside a large basin of attraction close to the origin
converge to the well studied H\'enon attractor, 
shown in Fig.~\ref{fig.henon}(a).
We let trajectories leak the system through the leak region~$I$: circle of radius~$\delta$ 
centered at some~($x_c,y_c$) on the attractor.
 The Sinai-Ruelle-Bowen measure
is the natural measure~$\mu$ describing the closed
system. 
All the initial densities discussed in Sec.~\ref{ssec.ic} 
are marked in Fig.~\ref{fig.henon}.
Panels (b)-(d) show the invariant saddle and its manifolds.
A new feature in comparison with saddles of naturally open systems is that the
unstable manifold is not a single fractal curve but consists of disjoint 
pieces cut by the leak and its images.

\begin{figure}[!bt]
\includegraphics[width=0.95\columnwidth]{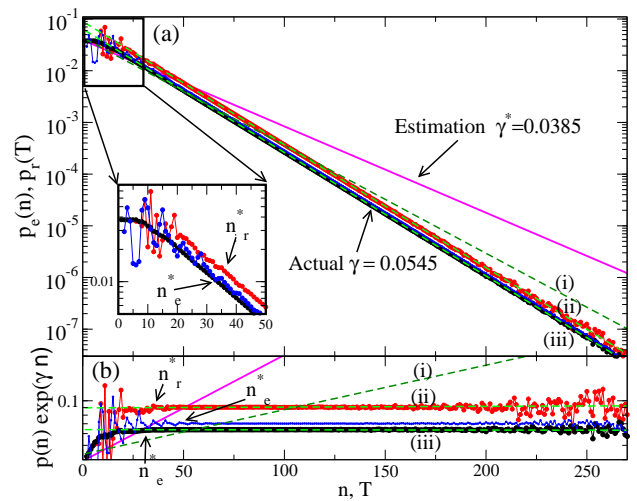}\\
\caption{ (Color online) (a) Escape time distributions for the H\'enon map~(\ref{eq.henon})
  with a leak of radius $\delta=0.05$ at~$(x_c,y_c)=(1.0804,0.0916)$. Results for initial densities~$\rho_r, \rho_\mu,$ and  $\rho_s$ (shown in Fig.~\ref{fig.henon}) are presented from top to bottom, respectively.
The thin solid line corresponds to $\gamma^*=0.0385$ given by~(\ref{eq.gammamu}) and strongly differs from the actual exponent~$\gamma=0.545$.
The inset shows the short time behavior.
(b) The same as in   (a) multiplied by $\exp(\gamma n)$. Three different 
procedures were employed to fit~$\gamma$ using $p(n)$ for $n<n^*$ (see
Sec.~\ref{sec.fitting}) and are indicated by the dashed lines (i)-(iii). In all
cases~$n^*=35$ and (i) corresponds to the use of relation~(\ref{eq.fit3}), 
 (ii) follows from relations~(\ref{eq.fitting.1}) and~(\ref{eq.fitting.2}) 
with the numerical value of $\langle n \rangle$ for~$\rho_0=\rho_r$, and (iii) from
relations~(\ref{eq.fitting.1}) and~(\ref{eq.fitting.2}) with the numerical value of
$\langle n \rangle$ for~$\rho_0=\rho_s$.  
}
\label{fig.etd}
\end{figure}

In Fig.~\ref{fig.etd} we show the escape time distribution for the initial densities {depicted} in Fig.~\ref{fig.henon}.
 It is apparent that all initial densities lead to {\em different} short time behavior but to the {\em
  same}  escape rate~$\gamma$, which is significantly different from the naive estimate~(\ref{eq.fpe}).

\vspace{0.5cm}

\begin{figure}[!bt]
\includegraphics[width=0.8\columnwidth]{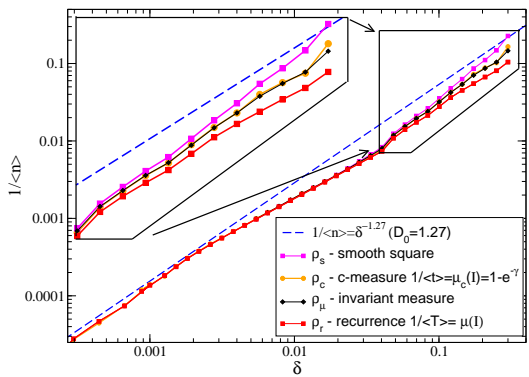}
\caption{ (Color online) Dependence of the inverse mean escape time 
$1/\langle n \rangle$
on the radius~$\delta$ of the leak for fixed~$(x_c,y_c)=(0.015714,0.268892)$. Initial densities equivalent
  to those shown in Fig.~\ref{fig.henon} are taken. The straight line
  corresponds to $\delta^{-D_0}$, where $D_0=1.27$ is the fractal dimension of 
  the attractor~\cite{alligood}. The inset shows a magnification of the main
  graph for large $\delta$.}   
\label{fig.tmedio}
\end{figure}

We investigate now the dependence of $\gamma$ and
$\mu(I)$ on the radius~$\delta$ of the leak. 
Since the natural and the c-measure have different fractal properties,
for small (but not infinitesimally small) $\delta$ 
one expects the scaling relation
\begin{equation}
\begin{array}{ll}
\frac{1}{\langle T \rangle} =\mu(I) \sim \delta^{D_1}, \\
\gamma \approx  \mu_c(I) \sim \delta^{1+d_s} = 
\delta^{1+(\lambda-\gamma)/\mid \lambda' \mid},
\end{array}
\end{equation}
where 
$D_1$ is the information dimension of the chaotic attractor, $d_s$ is
the partial information dimension of the saddle (system with leak)
along the stable direction (the c measure is smooth along the unstable 
one), 
and (\ref{eq.kantzgrassberger}) has been used.
The numerical comparison can be seen in
Fig.~\ref{fig.tmedio} where the inverse mean escape time is plotted for
different initial densities.
For the c-measure and $\delta \ll 1$,
$1/\langle n \rangle$ corresponds to $\mu_c(I)$ 
  through  relation~(\ref{eq.gammamuc}), while $1/\langle n \rangle=\mu(I)$
  with $\rho=\rho_r$. 
  The deviation between these two
  curves is a measure of 
  the error   of the naive estimate~(\ref{eq.fpe}). The general dependence,
  for $\delta \rightarrow 0$,
  is in agreement with the expected~$\delta^{-D_1}$ relation, by taking into account that
    $D_1 \lessapprox D_0$, where $D_0$ is the fractal dimension of the attractor. From the inset it is apparent that initial
    conditions $\rho_{\mu}$ and $\rho_s$  according to the natural measure 
    and to the homogeneous distribution, respectively 
    are closer to the results obtained with the density $\rho_c$ of the c-measure 
    than that obtained  with initial density $\rho_r$ 
    of the recurrence problem. In other words, 
    the mean escape time for typical smooth densities is better approximated by
    Eq.~(\ref{eq.kac3}) than by $1/\mu(I)$ [Eq.~(\ref{eq.kac})].

\begin{figure}[!bt]
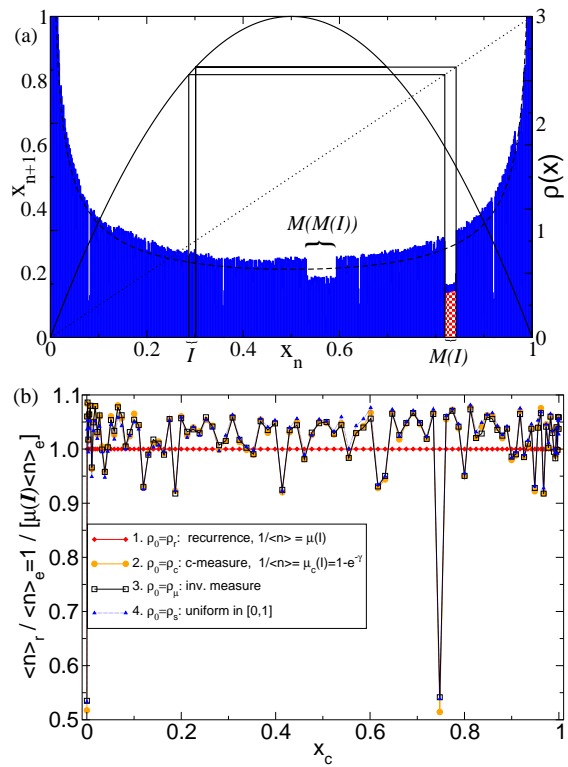

\includegraphics[width=0.85\columnwidth]{longo4a.eps}\\
\vspace{0.2cm}
\includegraphics[width=0.85\columnwidth]{longo4b.eps}
\caption{ (Color online) (a) Graphical representation of the logistic map (\ref{eq.logistic}) 
with a leak $I$ at $x_c=0.29427$ of half-width $\delta=0.00716$ so that $\mu(I)=0.01$. 
The dashed-line corresponds to $\rho_{\mu}(x)$, Eq.~(\ref{eq.logmeasure}), and the (blue)  histogram 
to a numerical distribution of the c-measure of the chaotic repeller 
obtained by leaking the system at~$I$.
The dotted (red) histogram corresponds to $\rho_r(x)$ given by
(\ref{eq.rhor1}) multiplied by $\mu(M(I))$. 
(b) Dependence of the inverse mean escape time, multiplied by $\langle n \rangle_r$, on the
leak position~$x_c$ in 
  the logistic map~(\ref{eq.logistic}) at an interval length 
such that $\mu(I)=0.01$. The initial densities used are (see Table~\ref{tab.ic}):
$\rho_r$: equivalent to recurrence, $\rho_{\mu}$: natural density
Eq.~(\ref{eq.logmeasure}),  and $\rho_s$: chosen uniform in  $0<x<1$. }
\label{fig.logistic}
\end{figure}

A systematic investigation of the dependence of $\gamma$ on the {\em position}
of the leak in the H\'enon map would be very involved. Apart from the difficulty
of having two dimensions of the phase space, the
natural (SRB) measure of the system is not known {\em a priori}. Therefore, in
order to fix the natural
measure of the leak $\mu(I)$, a different value of $\delta$ should be
chosen for each new $(x_c,y_c)$. We investigate therefore this issue
in the fully chaotic logistic map
\begin{equation}\label{eq.logistic}
x_{n+1}=4x_n(1-x_n). 
\end{equation}
In this case the density $\rho_{\mu}$ of the natural measure of the system is given by~\cite{ott}
\begin{equation}\label{eq.logmeasure}
\rho_\mu(x)={\pi}^{-1} {[x(1-x)]}^{-1/2}.
\end{equation}
The results are shown
in Fig.~\ref{fig.logistic}(b) and confirm the strong dependence of $\gamma$ on~$I$, as well as 
the conclusion drawn from Fig.~\ref{fig.tmedio} that for general smooth initial densities
$\gamma$ is better approximated by~(\ref{eq.kac3}) than by $1/\mu(I)$, reflecting once
more the fast convergence of $\rho_0$ to $\rho_c$.
Moreover, it is clear that for most~$x_c$ one has $\langle n
\rangle_e > \langle n \rangle_r$, i.e., $\mu(I) > \mu_c(I)$, in agreement with the
discussion at the end of Sec.~\ref{sec.rec}. The remarkable exceptions are related to the 
existence of short time periodic orbits inside the leak, as previously described in Refs.~\cite{paar,altmann,BY}. An explanation for these cases can be given as follows (see Sec.~\ref{sec.fitting} for details): if a periodic orbit of period $n_p$ is inside~$I$ we expect a high recurrence probability~$P(T=n_p)$; since $\langle T \rangle$ is fixed by~(\ref{eq.kac}), 
we predict that~$\gamma < \gamma^*$ and therefore $\langle n_r \rangle / \langle n_e \rangle < 1$, as observed in Fig.~\ref{fig.logistic}(b).

\section{nonhyperbolic Hamiltonian systems with power-law tails}\label{sec.standard}

\subsection{Theory}
Our discussion has covered
so far the case of strongly chaotic {systems} where
asymptotically in time, the distribution of escape and recurrence times is exponential
(for certain classes of systems this has been formally proved~\cite{hirata}). We
concentrate in this section on weakly chaotic systems where deviations from
exponential decay are observed for long times. 
The simplest examples of weakly chaotic systems are one-dimensional intermittent maps with
a marginal
fixed point, where an asymptotic power-law decay is proved \cite{PM,Moura}. 
The most prominent examples are, however, Hamiltonian systems with mixed phase space,
where asymptotically a power-law decay is observed.
This occurs in systems
possessing  isolated marginally unstable orbits~\cite{artuso}, sharply divided phase 
space~\cite{altmann.sharply}, 
hierarchical Kolmogorov-Arnold-Moser (KAM)
islands coexisting in a chaotic sea~\cite{CS,Zasl1,WHK,CK}, and also in
higher dimensions~\cite{higherD}. 
The power-law behavior is due to the nonhyperbolicity of the dynamics  and it is
present both for recurrences and escapes. The power-law
exponent is independent of the choice of~$I$ and~$\rho_0$, provided the latter is concentrated 
{\em away} from the nonhyperbolic regions. Below we show that the same holds for the intermediate time  exponential decay of the
recurrence and escape time distributions, for a given~$I$. 

The distribution of escape and recurrence times in Hamiltonian systems with
mixed phase space can thus be given as

\begin{equation}\label{eq.complete}
p(n) \approx \left\{ \begin{array}{ll}
  \text{ irregular } & \text{ for } n < n^*, \\
   \gamma a e^{-\gamma n} & \text{ for } n^* < n < n_{\alpha}, \\
   \gamma [a e^{-\gamma n} + b (\gamma n)^{-\alpha}] & \text{ for } n > n_{\alpha}, \\
\end{array} 
\right.
\end{equation}
where~$a e^{-\gamma n_\alpha}\gg b(\gamma n)^{-n_\alpha}$. The factor~$\gamma$ is written out on the right hand side in
order to facilitate the connection to the continuous time limit.
Contrary to what has previously been claimed ~\cite{alt}, $\gamma(a+b) \neq 1$ in 
Eq.~(\ref{eq.complete}).  

The question whether the asymptotic regime has a well defined (universal)
power-law is still under investigation for the case of area-preserving maps (see Ref.~\cite{CK} for the latest result 
that indicates that $\alpha \approx 2.57$). Here, for simplicity we write the asymptotic decay as $n^{-\alpha}$, but 
it is meant to describe the power-law like behavior usually observed. 

We define the cross-over time $n_c$ between exponential and 
power-law  decay as:
\begin{equation}\label{eq.defnc}
 a e^{-\gamma n_c}= b (\gamma n_c)^{-\alpha} \Rightarrow p(n_c)= 2 \gamma a e^{-\gamma  n_c}.
\end{equation}

In the framework of {\em recurrences}, the different times in
Eqs.~(\ref{eq.complete}) and (\ref{eq.defnc}) can be interpreted as:

\begin{itemize}

\item[$n^*$:] is the short-time memory, within which fluctuations due to short time
  periodic orbits appear. After this time the recurrences loose correlation 
  and a random approximation for the return is valid (Poisson process).

\item[$n_{\alpha}$:] is the minimum time a trajectory inside~$I$ takes to approach
  the nonhyperbolic region and return to $I$.

\item[$n_c$:] is the time when returning without
  touching the nonhyperbolic region becomes as probable as returning after touching. 

\end{itemize}
If $I$ is small and away from any nonhyperbolic region we expect
$n^*<n_\alpha<n_c$.

In terms of {\em escapes}, the times in Eq.~(\ref{eq.complete}) are interpreted
as:
\begin{itemize}

\item[$n^*$:] is a convergence time which is proportional to~$1/|\lambda'|$,
  where~$\lambda'$ is the  negative Lyapunov exponent of the saddle (the time to relax
  to the hyperbolic component of the saddle along its stable manifold).

\item[$n_{\alpha}$:] is the time needed to approach  the nonhyperbolic part of the saddle.

\item[$n_c$:] is the time when the nonhyperbolic part of the saddle becomes
  as important as the hyperbolic one.

\end{itemize}
In this case, the key assumption $n^*<n_\alpha$ is related to the initial
density~$\rho_0$ and leak $I$ being away from the sticky regions. Being invariant, the
division of the saddle is of course independent of the initial density. However,  for initial conditions
touching the KAM island the importance of the exponential decay can be made arbitrarily
small (see Fig.~\ref{fig.7} below) and the power-law exponent in~(\ref{eq.complete}) is
modified to~$\alpha'=\alpha-1$, see Ref.~\cite{pikovsky,altmann.sharply}
(see also Ref.~\cite{hirata}).  

The above interpretation of $p(n)$ given by~(\ref{eq.complete}) 
expresses the view, first suggested in \cite{JTZ}, that the
nonhyperbolic component of the saddle is {\em approached through the
  hyperbolic part} \cite{JTZ,alt,ML2,LFO}, i.e., the fraction of trajectories that reach the
nonhyperbolic part is proportional to those that reach the hyperbolic one. When the size of the leak~$I$ decreases, the hyperbolic component 
increases while the nonhyperbolic one remains the same, meaning that the
above picture is even more accurate. In this case we can expect that the {fraction of
  trajectories that arrive at the nonhyperbolic component is proportional to the fraction
  of trajectories that approach the hyperbolic component and that therefore the}
  ratio~$b/a$ [see Eq.~(\ref{eq.complete})] has a
weak dependence on~$\gamma$. 
Introducing this assumption in the definition of the cross-over time~$n_c$ [Eq.~(\ref{eq.defnc})] we obtain 
\begin{equation}\label{eq.nc}
n_c \sim 1/\gamma  \;\; [ \approx 1/\mu(I)\;\; \text{ for small } \;\; \mu(I)],
\end{equation} 
meaning that $n_c$ is proportional to the reciprocal of the escape rate~$\gamma$. The proportionality 
constant depends on the size of the nonhyperbolic regions. {We have verified} relation~(\ref{eq.nc}) 
for different maps, what emphasizes once more that for small~$\gamma$ a well defined exponential decay exists for intermediate times~$n^* < n < n_c$. 

\subsection{Numerical results for the standard map}

\begin{figure}[!bt]
\includegraphics[width=0.9\columnwidth]{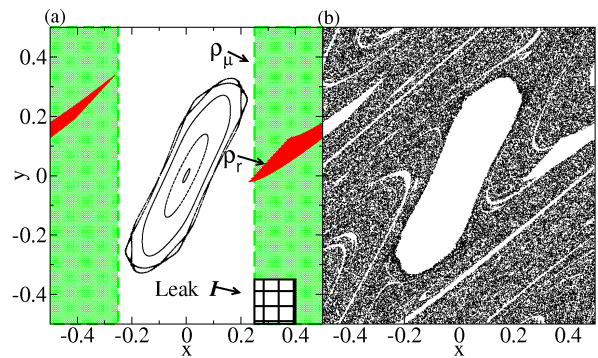}
\caption{ (Color online) (a) Phase space of the standard map~(\ref{eq.standard}) with
   $I=[0.25<x<0.25+\varepsilon,-0.5<y<-0.5+\varepsilon]$ of
   size~$\varepsilon=0.15$, indicated by a checkered box. The 
  KAM island (nonhyperbolic part of the saddle) is in the center of the figure. 
     The support of~$\rho_r$ 
at  $M(I)$ is marked as a
dark (red) region. (b) The unstable manifold of the 
  hyperbolic part of the saddle ( black dots), obtained by using the method
   of~\cite{TG} choosing an escape time~$n_s=80 < n_c=307$.
   The support of $\rho_r$ 
   is fully outside the unstable manifold.
   }
\label{fig.5}
\end{figure}

We illustrate the previous results in the standard map~\cite{CS,Zasl2}:
\begin{equation}\label{eq.standard}
y_{n+1} = y_n-0.52 \sin(2\pi x_n),\;\;\;\;\;
x_{n+1} =x_n+y_{n+1}.
\end{equation}
The phase space of this map is shown in Fig.~\ref{fig.5}. Chaotic trajectories stick for
an algebraic long time to the hierarchical border of the KAM island, shown in the center
of Fig.~\ref{fig.5}(a), that constitutes the nonhyperbolic component of the saddle. The
exponential decay is governed by the hyperbolic component of the saddle, whose
  unstable manifold is shown in Fig.~\ref{fig.5}(b)  for a specific leak~$I$.  
This unstable manifold corresponds to the support of the conditionally invariant density~$\rho_c$.
Notice its apparent filamentary character and that it is present inside~$I$,
{but does not enter $M(I)$. }

\begin{figure}[!bt]
\includegraphics[width=0.9\columnwidth]{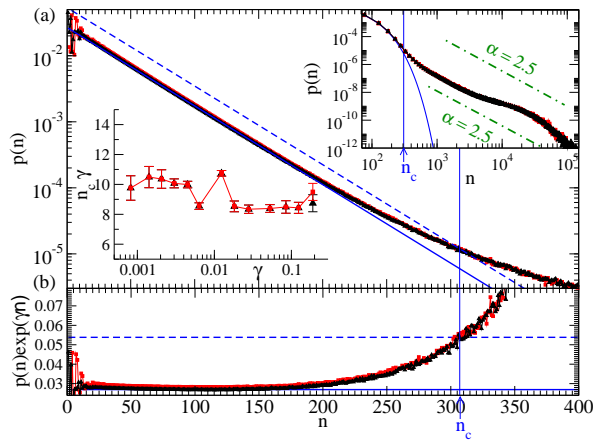}
\caption{ (Color online) (a) Escape time distributions in the
standard map with $\varepsilon=0.15$, as shown in Fig.~\ref{fig.5}. 
Results are shown for $\rho_0=\rho_r$ (squares/below)
and for $\rho_0=\rho_\mu$ in $|x|>0.25$. 
Lower inset: scaling of $n_c \gamma$ with~$\gamma$
obtained by changing $\varepsilon$ in~$0.03\le \varepsilon\le 0.4$. 
Upper inset: 
log-log plot of the main graph over a long period of time.
(b) $p_{e,r}$ multiplied by $\exp(\gamma n)$, where
$\gamma=0.027403$. 
}
\label{fig.6}
\end{figure}

\begin{figure}[!bt]
\includegraphics[width=1\columnwidth]{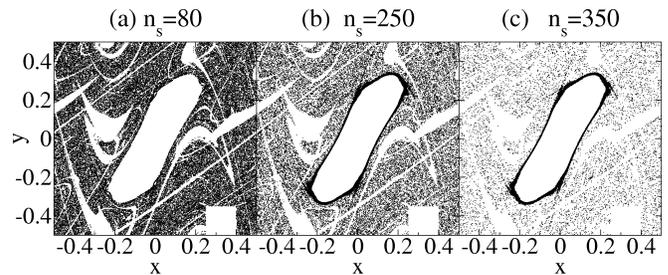}
\caption{ The saddle characterizing the dynamics at different times, $n_s$, given on top of the panels,
obtained by the method of Ref.~\cite{TG}. For $n_s<n_c=307$ in (a), the hyperbolic component obtained is clearly disjoint from he region of the KAM island. By increasing $n_s$ as in (b) and (c) the nonhyperbolic component appears with an increasing weight.}
\label{fig.7}
\end{figure}

The distribution of escape and recurrence time for the system presented in Fig.~\ref{fig.5} is shown in Fig.~\ref{fig.6}. As in the case of the H\'enon map (Fig.~\ref{fig.etd}), short time oscillations and an exponential decay are clearly visible.
However, in this case a slower decay is observed for times $n>n_c\approx 307$. In the upper inset of Fig.~\ref{fig.6}(a),
 a log-log plot of the main graph shows that the long-time decay is roughly a power-law with~$\alpha \approx2.5$. 
We have also obtained numerically the value of the cross-over time~$n_c$ for leak/recurrence regions~$I$ of different sizes by changing~$\varepsilon$ defined in Fig.~\ref{fig.5}. The results are shown in the lower inset of Fig.~\ref{fig.6}(a), and support the theoretical scaling~(\ref{eq.nc}). Remarkably, the value 
of~$n_c$ is almost identical for recurrence and escape times, the same being valid also for other initial distributions~$\rho_0$ away form the KAM island. 
This provides additional support to the picture that~$\rho_0$ quickly converges to the
hyperbolic component of the saddle, and that the nonhyperbolic component is approached in
an universal way through the hyperbolic component. 
The transition between the hyperbolic
and nonhyperbolic components of the saddle is illustrated in Fig.~\ref{fig.7}.

\section{Fitting procedure to obtain the escape rate}\label{sec.fitting}
In the previous sections we have seen that the escape rate~$\gamma$ can be
formally written as a function of the c-measure~$\mu_c(I)$
of the leak region~$I$ through relation~(\ref{eq.gammamuc}). However,
this relation is of little practical use if one wants to
determine~$\gamma$. The reason is that usually only the Lebesgue measure is
known a priori and the c-measure $\mu_c$ may 
change dramatically depending on the position of the leak. Its numerical
calculation is typically much more involved than the determination of~$\gamma$
itself. In this section we search for an efficient method to determine~$\gamma$
that goes beyond the rough naive estimate~$-\ln(1-\mu(I))$, given by relations~(\ref{eq.fpe}) and~(\ref{eq.gammamu}).

In general open hyperbolic systems a dense set of periodic orbits exists
embedded into the nonattracting chaotic set. This can be used to effectively obtain~$\gamma$ as~\cite{ott,dorfman} 
\begin{equation}\label{eq.upo}
e^{-n\gamma}=\sum_{\Gamma_{i,n}} \prod |\Lambda (\Gamma_{i,n})|,
\end{equation}
where the product is over the expanding eigenvalues~$\Lambda$ of each periodic
point $\Gamma_{i,n}$
belonging to the same periodic orbit of length $n$.
The sum runs over all periodic
orbits of period~$n$ that remain inside the system, i.e., {\em outside} the
leak region~$I$. 

We want to take further advantage here of the possibility of closing the leak of the
system. Note that in the closed system the decay exponent is~$\gamma=0$ and the sum in
Eq.~(\ref{eq.upo}) runs over all periodic orbits (inside and outside the
leak region). We can therefore give $\gamma$ in terms of the periodic orbits {\em inside} the leak region as 
\begin{equation}\label{eq.upos}
 1-e^{-n\gamma}=\sum_{\Gamma_{i,n} inside} \prod |\Lambda (\Gamma_{i,n})|,
\end{equation}
From this expression, it is clear that short time periodic orbits have a strong influence
on~$\gamma$ since not only the primitive orbits, but also all their multiples appear in the sum. This dependence is apparent in Fig.~\ref{fig.logistic}(b) and was previously reported in Refs.~\cite{paar,altmann,BY}.
The application of~(\ref{eq.upos}) requires, however,  again the identification of the unstable
periodic orbits of high periods. Since the deviations of $\gamma$ from the
estimate~$\gamma^*=-\ln(1-\mu(I))$ has been related to {\em short} time periodic orbits we
can expect to determine $\gamma$ from these orbits. A method for this purpose
was described in Ref.~\cite{altmann} in the context of recurrences. We adapt it here in the context of escape. 
 
The distribution of escape times is written as in Eq.~(\ref{eq.pesc}). The oscillatory part
for times~$n<n^*$ 
will be denoted by $p_0(n)$ and the remaining exponential
part by~$p_{exp}(n)\equiv g \exp{(-\gamma n)}$. The two parameters~$g$ and $\gamma$ of the asymptotic
exponential decay can be obtained then by imposing the following two conditions:
normalization  
\begin{equation}\label{eq.fitting.1}
\sum_{n=1}^{n^{*}-1} p_{0}(n)+\sum_{n=n^{*}}^{\infty} p_{exp}(n)=1\,,
\end{equation}
and the value of the mean 
\begin{equation}\label{eq.fitting.2}
\langle n \rangle=\sum_{n=1}^{n^{*}-1} n p_{0}(n)+\sum_{n=n^{*}}^{\infty} n p_{exp}(n)=\dfrac{1}{\mu_i(I)}\,.
\end{equation}
For recurrences, or $\rho_0=\rho_r$, we can use Kac's lemma~(\ref{eq.kac}) and thus~$\mu_i(I)=\mu(I)$, the
natural measure of the leak. On the other hand, when initial densities are
proportional to the natural measure in the escape problem, the best procedure
is to replace~(\ref{eq.kac}) by~(\ref{eq.kac3}) and therefore
use~$\mu_i(I)=\mu_c(I)=1-e^{-\gamma}$ on the right hand side of Eq.~(\ref{eq.fitting.2}). In the last
case, we can solve Eqs.~(\ref{eq.fitting.1}) and~(\ref{eq.fitting.2}) in terms
of~$\gamma$, which is expressed as the solution of
\begin{equation}\label{eq.fit3}
(1-S_1) \frac{n^*e^\gamma+1-n^*}{e^\gamma-1}=\frac{1}{1-e^{-\gamma}}-S_2,
\end{equation}
where~$S_1=\sum_{n=1}^{n^{*}-1} p_{0}(n)$ and~$S_2=\sum_{n=1}^{n^{*}-1} n p_{0}(n)$. Both
fitting procedures, the one given by Eqs.~(\ref{eq.fitting.1})-(\ref{eq.fitting.2}) and
the one using Eq.~(\ref{eq.fit3}), were employed for the case of the H\'enon map and are
marked as dotted lines in Fig.~\ref{fig.etd}. Notice that the fitting obtained using
Eq.~(\ref{eq.fit3}) alone is worse than the one obtained using as additional information
the numerical value of~$\langle n_e \rangle$  in
Eqs~(\ref{eq.fitting.1})-(\ref{eq.fitting.2}), but better than the random 
estimate  (\ref{eq.gammamu}). 

The main advantage of this procedure is that, even if theoretically large~$n^*$ in
Eqs.~(\ref{eq.fitting.1})-(\ref{eq.fit3}) would render better results, in 
practice the value $n^*$ is fairly small 
as numerical experience shows ($n^* \lessapprox \langle n \rangle$)
The values of 
$p(n)$ for $n<n^*$ are thus obtained without much computational effort already
with a high precision but allow the computation of the asymptotic
decay~$\gamma$. Moreover, this shows that important information about the
long-time dynamics is already contained in the short time dynamics.

In the nonhyperbolic case one has to take into account also the contribution of the
power-law tail. 
In the continuous limit, the following contributions:
\begin{equation}
\begin{array}{ll}
S_3= \sum_{n=n_\alpha}^\infty \gamma b(\gamma n)^{-\alpha} =  b (\gamma n_\alpha)^{1-\alpha} / (\alpha-1), \\ 
S_4= \sum_{n=n_\alpha}^\infty \gamma n b(\gamma n)^{-\alpha} = b \gamma^{1-\alpha} n_\alpha^{2-\alpha} / (\alpha-2),
\end{array}
\end{equation}
are to be added as new terms to the left hand side of Eqs.~(\ref{eq.fitting.1})
and~(\ref{eq.fitting.2}), respectively.
Since $n_\alpha \sim n_c \sim 1/\mu(I)$, and $\alpha>2$,  one sees that these terms become negligible for
small recurrence/leak regions.


\section{Conclusions}\label{sec.conclusion}

Experimental and observational measurements {often} occur through holes or leaks that naturally 
exist or are deliberately introduced in an otherwise closed dynamical 
system~\cite{PY,BGOB,LS,DS,paar,pierrehumbert,schneider,SHA,nagler.astro,motter}.  
The relevant observable quantity in such systems with leaks is the distribution of escape
times from inside the system. In strongly chaotic systems, this distributions decays
exponentially. 
Applying the ergodic theory of transient chaos \cite{PY} to chaotic systems with leaks an
expression for the 
escape rate~$\gamma_e$ [Eq.~(\ref{eq.gammamuc})] is obtained~\cite{paar} in terms of the c-measure (associated to the invariant chaotic saddle) of the leak~$I$. 
 We have  provided a theoretical framework to understand the strong dependence of~$\gamma$ and~$\langle n_e \rangle$ on the leak size (Fig.~\ref{fig.tmedio}) and position~(Fig.~\ref{fig.logistic}).  
In the (unrealistic) limit of small leak, the expression of the escape rate converges to a previously known relation 
[Eqs.~(\ref{eq.smallmu})], based on the invariant measure of the closed system. 
Altogether, our results help to understand previous
observations~\cite{paar,schneider,altmann,BKG} and set a theoretical framework for
future applications. 

We have shown that the classical problem of Poincar\'e recurrences in closed systems can be described as a problem of
 escape from a system with a leak, once  the recurrence region is identified with the leak~$I$ and the 
initial density is chosen properly. This allows us to treat both problems in an unified framework and 
compare previous similar results that were reported independently in both fields
(compare Refs.~\cite{paar,schneider} and~\cite{altmann}), and to adapt (in
Sec.~\ref{sec.fitting}) a previous method to efficiently obtain~$\gamma_e$. More
surprisingly, we show that the relaxation rate of the distribution of Poincar\'e
recurrences~$\gamma_r$ is equal to~$\gamma_e$~[Eq.~(\ref{eq.gamma})] but that the mean
recurrence time $\langle n_r \rangle$ given  
by Kac's lemma~(\ref{eq.kac}) differs from the typical mean escape time [the analog of Kac's lemma for leaked
systems is~(\ref{eq.kac2})].  These results are summarized in Table~\ref{tab.ic} and
provide a more detailed account of the results published in Ref.~\cite{letter}.

In weakly chaotic systems the asymptotic exponential decay is replaced by a power-law decay. We have shown, however,
that if the leak region and the initial distribution are away from nonhyperbolic regions, an exponential decay 
is still well defined for intermediate times. We have obtained the scaling of the cross-over time~$n_c$ between the
two regimes [Eq.~(\ref{eq.nc})] which confirms the importance of the exponential decay for small leaks. 
Altogether, these results justify an effective splitting of the chaotic saddle in a 
nonhyperbolic component, related to the asymptotic power-law decay, and an hyperbolic component, related to the 
intermediate exponential decay for which our previous results apply. This splitting is particularly important in 
the case of Hamiltonian systems (found, e.g., in fluid dynamics or optical applications) where the generic phase space
shows a mixture of regions of regular and chaotic motion.

\begin{figure}[t]
\includegraphics[width=1\columnwidth]{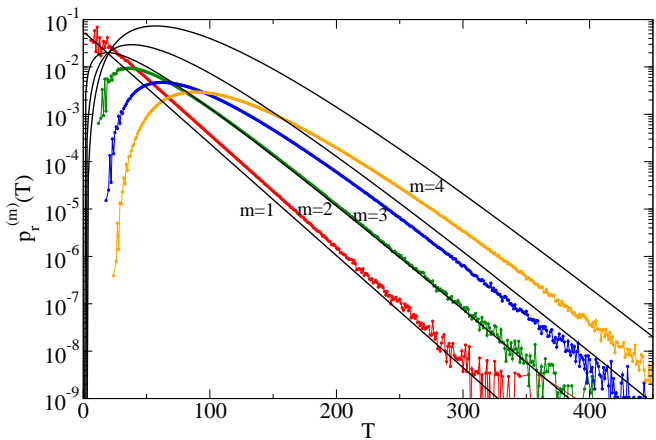}
\caption{ (Color online) Distribution of the $m$-th recurrence times ($m=1,...,4$, from bottom to top) in
  the H\'enon map with parameters as in Fig.~\ref{fig.etd}. The solid lines correspond to
  the binomial distribution~(\ref{m}) using $\mu=
\mu_c(I)$.
For long times, the same decay rate~$\gamma=\ln{(1-\mu_c(I))}$ is observed in all curves. The displacement between
  the distribution and the simulation, as in the~$m=1$ case
discussed previously, can be
  related to the short time oscillations that modify the pre-factors of the distribution.
}
\label{fig.8}
\end{figure}

Finally, it is worth  considering briefly two extensions of the previous results that have potentially
interesting applications. 
First, consider the case when the leak consists of two or more disjoint components (of
relevance, e.g., for the problem of resetting in hydrodynamical flows~\cite{pierrehumbert}). 
All our results apply to such cases as well, with the remark that the escape rate is not the mere 
sum of the escape rates characterizing the components, due to the overlap among pre-images,
as pointed out by Buljan and Paar~\cite{paar} and Bunimovich and Dettmann~\cite{BD}.   
A second extension is to consider the distribution of the second, third, \ldots, $m$-th recurrence
times. In the notation of Sec.~\ref{sec.recurrence}, the $m$-th recurrence time is defined 
as  $T^{(m)}_i\equiv n_i-n_{i-r}$. In view of the analogy to escape presented in
Sec.~\ref{sec.rec}, this corresponds to the distribution of escape times taking place only
after entering the leak~$I$ for the $m$-th time. The naive (binomial)
approximation~(\ref{eq.binomial}) can be extended to this case as~\cite{altmann} 
\begin{equation}
p_r^{(m)}(T) = \frac{(T-1)!}{(T-m)!(m-1)!} {\mu(I)}^m {(1-\mu(I))}^{T-m},
\label{m}
\end{equation}
and implies that the asymptotic decay  
rate  $\gamma^{*(m)} = \ln{(1-\mu(I))}$ [see (\ref{eq.gammamu})] remains valid for any $m$. 
In Fig.~\ref{fig.8} we show numerical results for the H\'enon map. For $m>1$ the
distributions take, of course, the value zero for $n=1$, start growing, 
and then after a maximum is reached, an exponential decay 
sets in. The decay rate has been found again to be different from the binomial
approximation, but apparently also {\em independent of $m$}, i.e., to have the value 
$$\gamma^{(m)} = \gamma= \ln{(1-\mu_c(I))}.$$
discussed in the context of the first recurrence times  [see Eq.~(\ref{eq.fpe})]. 
To understand this result, let us first compare
the invariant saddles for~$m=1$ ($S_1$) and $m=2$ ($S_2$). 
It is easy to see that $S_1\subset S_2$. Trajectories that belong to $S_2$ but {\em not} to~$S_1$ 
 must have a point $P$ inside $I$ that never returns to $I$ again.
Hence, these
trajectories necessarily approach $S_1$ for both~$t \rightarrow \pm\infty$.
To be outside $I$ for
$t \rightarrow \infty$, $P$ has to belong to the unstable manifold
of $S_1$.
To be outside $I$ for
$t \rightarrow -\infty$, the image point $M(P)$ taken with respect to the
closed system's map
has to be on the stable manifold of $S_1$.
Therefore,  points $P$ belong to the
intersection of the unstable manifold of~$S_1$ and the pre-image (using map~$M$) of the stable manifold
of~$S_1$ and also approach~$S_1$ asymptotically. 
Accordingly, trajectories escaping for long times in the case~$m=2$ appear still to be governed by S1 having
therefore the escape rate $\gamma$. 
This argument can be extended straightforwardly to any finite ~$m$ and can also be important for the
the case of partial escape in optical systems~\cite{altmann.optics}.

\acknowledgments
We are indebted to J. Br\"ocker, L.A. Bunimovich,  G. Gy\"orgyi, H. Kantz, G. Del Magno, A. E. Motter,
and A. Pikovsky for useful discussions. We thank the anonymous referees for their
valuable inputs. This research was supported by the OTKA grants T47233, T72037.


\end{document}